\documentclass[12pt]{iopart}
\usepackage{graphicx}
\usepackage{color}
\begin{document}

\title{Flux-free conductance modulation in a helical Aharonov-Bohm interferometer}

\author{Hisao Taira$^{\rm 1,2,3)}$ and Hiroyuki Shima$^{\rm 1,2)}$}

\address{${}^{\rm 1)}$ Department of Applied Physics, Graduate School of 
Engineering, Hokkaido University, Sapporo 060-8628, Japan }

\address{${}^{\rm 2)}$ Department of Applied Mathematics 3, 
LaC${\rm \grave{a}}$N, Universitat Polit${\rm \grave{e}}$cnica de Catalunya 
(UPC), Barcelona 08034, Spain}

\address{${}^{\rm 3)}$ Department of Physics, 
The Chinese University of Hong Kong (CUHK), Shatin, New Territories, 
Hong Kong }

\ead{taira@eng.hokudai.ac.jp}
\begin{abstract}
A novel conductance oscillation in a twisted quantum ring composed of 
a helical atomic configuration is theoretically predicted. 
Internal torsion of the ring is found to cause a quantum phase shift 
in the wavefunction that describes the electron's motion 
along the ring. 
The resulting conductance oscillation is free from magnetic flux 
penetrating inside the ring, which is in complete contrast with 
the ordinary Aharonov-Bohm effect observed in untwisted quantum rings. 
\end{abstract}

\maketitle

\section{Introduction}

The Aharonov-Bohm (AB) effect is a pivotal manifestation of geometric phase 
governing quantum dynamics \cite{4}.
It occurs when a charged particle travels along a coherent loop threaded by 
external magnetic flux;
the particle's wave function acquires an additional quantum phase
that influences an interference pattern.
Experimental efforts to confirm its topological nature \cite{2,5}
as well as possible applications toward quantum computations \cite{1,1a}
have made it a topic of more broader interest than ever 
\cite{jpcm,cano}.

The AB effect was originally predicted for charged particles 
moving around magnetic flux.
Since then, it has been generalized to neutral particles having magnetic 
\cite{6,9,10}
or electric \cite{7,8} dipole momenta that travel around a line of electric 
or magnetic charges, respectively.
A unified picture for the three phenomena was established 
in the framework of the electromagnetic duality, 
which further predicted another interference phenomenon called the 
dual AB effect \cite{11}.
It is noteworthy that the AB effect has many analogues; 
in fact, light penetrating through an optical medium \cite{13}, 
quasiparticles moving in Bose-Einstein condensates \cite{14,14a,14b,14c},
and particles in a gravitational background \cite{15a,18,prd} have been 
suggested to exhibit AB-like phenomena. 
In addition, recent attempts to reveal the Dirac fermion dynamics 
in the AB ring 
\cite{jpcm2} and to unveil the ponderomotive AB effect driven by 
laser pulses \cite{njp} are quite intriguing. 
The series of work evidences the relevance of the effect to 
diverse fields in physics. 


In the present paper, we propose a distinct class of the AB-like 
interference effect 
for non-interacting charged particles that moves along a helical circuit, i.e., 
a quantum ring consisting of helical atomic structure. 
Surprisingly, the effect requires no magnetic flux threading 
inside the ring, which is in complete contrast to the ordinary 
AB effect. 
Such the flux-free interference effect originates from a
torsion-induced vector potential $\mbox{\boldmath $A$}_{\rm eff}$ that appears 
in the effective Hamiltonian describing the particle's motion 
along the helical circuit \cite{torsion1,torsion2,entin}; 
a similar torsion-induced 
effect was found in twisted optical waveguides \cite{longhi}. 
We demonstrate that an additional phase associated with 
$\mbox{\boldmath $A$}_{\rm eff}$
results in a conductance oscillation whose pattern 
is determined by the helicity of the atomic configuration.

\section{Basic equation and electronic eigenstates}

We consider a twisted quantum ring that has the ring radius $R_1$ and 
a uniform circular cross section with the tube radius $R_2$ $(\ll R_1)$.
The ring is composed of a helical atomic configuration 
around the centroidal axis $C$ parametrized by $q_0$. 
Using an appropriate reference frame $(q_0, q_1, q_2)$, a point 
in the vicinity of $C$ is represented by
$\mbox{\boldmath $R$}=\mbox{\boldmath $r$}(q_0)+q_1\mbox{\boldmath $e$}_1(q_0)+q_2\mbox{\boldmath $e$}_2(q_0)$, 
where the set ($\mbox{\boldmath $e$}_0,\mbox{\boldmath $e$}_1,
\mbox{\boldmath $e$}_2$) with $\mbox{\boldmath $e$}_0 \equiv \partial_0 
\mbox{\boldmath $R$}$ and $|\mbox{\boldmath $e$}_1|=|\mbox{\boldmath $e$}_2|=1$
forms a right-handed orthogonal triad and
$\partial_a \equiv \partial/\partial q_a \hspace{1.0mm} (a=0,1,2)$. 
The vectors $\mbox{\boldmath $e$}_1$ and $\mbox{\boldmath $e$}_2$ rotate
along $C$ with the same rotation rate as that of helical atomic configuration,
as a result of which the torsion, 
$\tau \equiv \mbox{\boldmath $e$}_2 \cdot \partial_0\mbox{\boldmath $e$}_1$, 
of the reference frame equals the internal torsion of the atomic 
configuration. 
Such the twisted reference frame can be useful for analyzing physical 
properties of actual twisted nanowires that were experimentally fabricated 
\cite{nature,prl} or theoretically suggested 
\cite{dandoloff,dafonseca,prl2,prb,Mladenov}. 

Using the twisted reference frame, 
the motion of non-interacting electrons is written by the Schr\"odinger equation such as \cite{Jensen,Dacosta}
\begin{eqnarray}
\mu \sum_{a,b=0}^{2}\frac{1}{\sqrt{g}}\partial_a
\left(\sqrt{g}g^{ab}\partial_b \right)\phi+V\phi=E\phi,
\label{eq:syure1}
\end{eqnarray}
where $\mu \equiv -\hbar^2/(2m^*)$ with an effective mass $m^*$ and 
\begin{eqnarray}
g={\rm det} [g_{ab}], \ \ g_{ab} = \partial_a\bi{R}\cdot\partial_b \bi{R},\ \ g^{ab}=g_{ab}^{-1}. \ \ [a,b=0,1,2]
\label{eq:metric}
\end{eqnarray}
In equation (\ref{eq:syure1}), $V$ represents a potential that confines  the transverse motion of 
electrons within the cross section. 
Components of the tensor $g_{ab}$ in equation (\ref{eq:metric}) are 
given by
\begin{eqnarray}
g_{00}&=&\gamma^4+\tau^2\left(q_1^2+q_2^2\right), \ \ g_{01}=g_{10}=-\tau q_2, \ \ g_{02}=g_{20}=\tau q_1,\nonumber \\
g_{ij}&=&\delta_{ij},
\hspace{1.0mm} [i,j=1,2],
\label{eq:metric2}
\end{eqnarray}
where $\displaystyle \gamma=\sum_{i=1}^2(1-\kappa_iq_i)^{1/2}$ and $\kappa_i = \mbox{\boldmath $e$}_0 \cdot \partial_0 \mbox{\boldmath $e$}_i$. 
The elements $g^{ab}$ are those of the $3\times3$ matrix $[g^{ab}]$ 
inverse to $[g_{ab}]$, and thus they can read as 
\begin{eqnarray}
g^{00}&=&\gamma^{-4}, \ \  g^{01}=g^{10}=\gamma^{-4}\tau q_2, \ \  
g^{02}=g^{20}= -\gamma^{-4}\tau q_1, \nonumber \\
g^{ij}&=&\delta_{ij}+\gamma^{-4}\tau^2\left[ 
\left( q_1^2+q_2^2\right)\delta_{ij}-q_i q_j \right].
\label{eq:metric3}
\end{eqnarray}

For simplicity, we assume that geometric deviation of the twisted 
frame from the orthogonal one is sufficiently smooth and small so that 
$(\kappa_1^2+\kappa_2^2)^{1/2}R_2 \ll1$ and $\tau R_2 \ll 1$. 
These assumptions allow to obtain \cite{torsion1,taira2}
\begin{figure}
\begin{center}
\includegraphics[width=8cm]{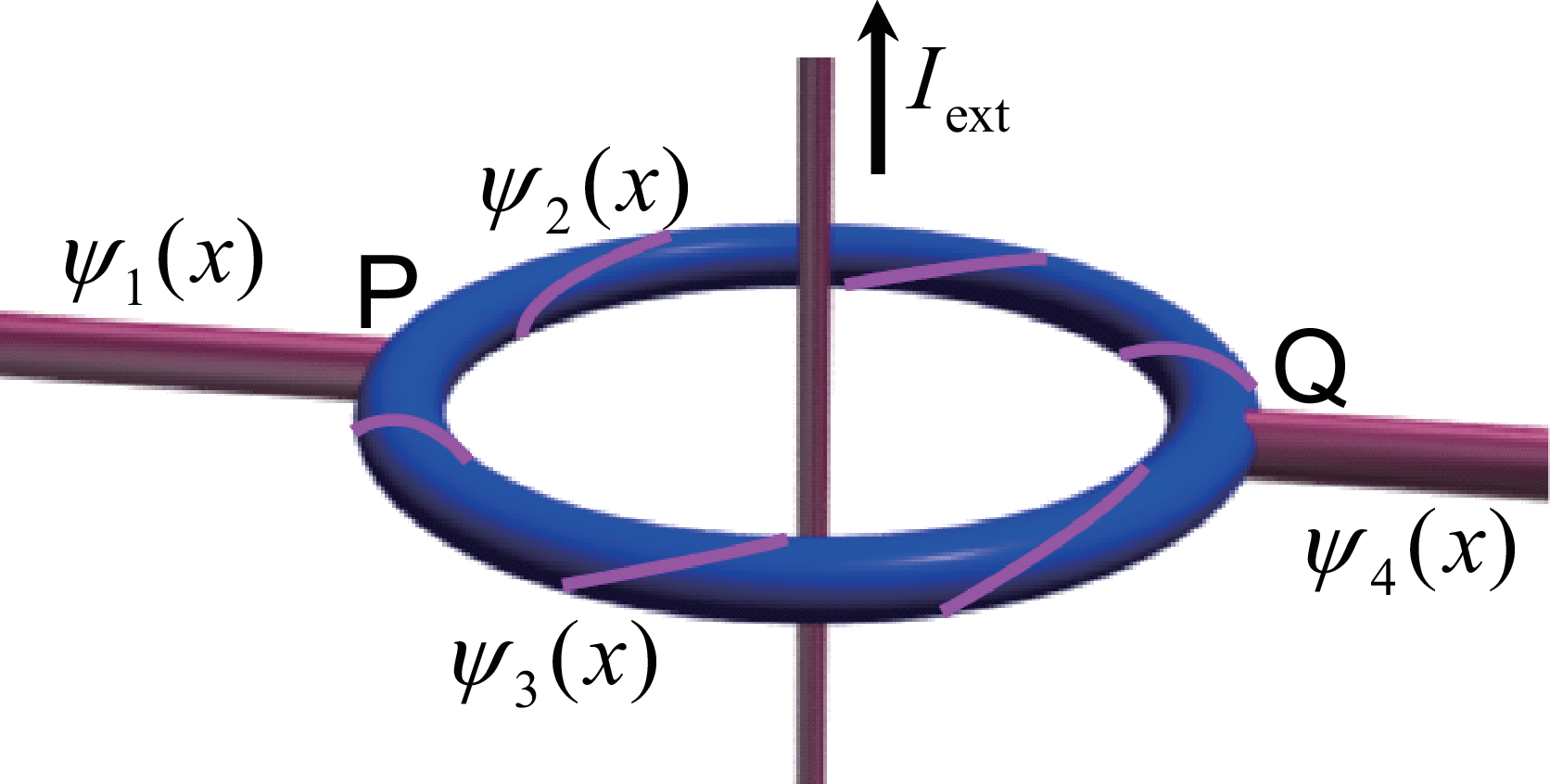}
\caption{\label{fig:fig1} 
Two terminal electron interferometer 
based on a twisted quantum ring encircling external 
current flow $I_{\rm ext}$. 
Each wavefunction $\psi_s(x)\ (s=1,2,3,4)$ describes the electron's 
motion lying at the lead ($s=1,4$) or the arc ($s=2,3$) as 
indicated in the figure.
}
\end{center}
\end{figure}
%
\begin{eqnarray}
\mu \left[\partial_1^2+\partial_2^2+\left( \partial_0-\frac{i\tau L}{\hbar} \right)^2
+\frac{1}{4R_1^2} \right] \phi +V\phi = E \phi,
\label{eq:syure2}
\end{eqnarray}
where 
$L\equiv -i\hbar(q_1 \partial_2 - q_2 \partial_1)$ is the angular 
momentum operator in the cross section. 
Eigenfunctions of the equation (\ref{eq:syure2}) have the form
\begin{eqnarray}
\phi(q_0, q_1, q_2)=\psi(q_0) \sum^N_{n=1} c_n u_n(q_1, q_2). 
\label{eq:separationvariables}
\end{eqnarray}
Here, $u_j(q_1,q_2)$ are $N$-fold eigenfunctions in the cross section 
\footnote{The value of $N$ depends on the eigenenergy $E$ of 
equation (\ref{eq:syure2}) to which $\phi$ belongs, and also on the 
shape of the cross section in general.} and $\psi(q_0)$ describes the axial motion of electrons along the 
twisted wire.
It follows from equations (\ref{eq:syure2}) and (\ref{eq:separationvariables}) that $\psi(q_0)$ obeys the effective one-dimensional 
Schr\"odinger equation such as 
\begin{eqnarray}
\mu \left[\left( \partial_0-\frac{i\tau \langle L \rangle}{\hbar} \right)^2
+\frac{1}{4R_1^2} - \frac{\tau^2}{\hbar^2} \left( \langle L^2 \rangle - \langle L \rangle^2 \right)\right] \psi(q_0) = \epsilon \psi(q_0),
\label{eq:1dsyure2}
\end{eqnarray}
The angular brackets $\langle \cdots \rangle$ in equation (\ref{eq:1dsyure2}) 
indicate to take an expectation value 
with respect to the two-dimensional ground-state function in the 
cross section. 
We see from equation (\ref{eq:1dsyure2}) that the product 
$\tau \langle L \rangle/\hbar$ plays a role of the effective vector potential 
$\mbox{\boldmath $A$}_{\rm eff}$ we have mentioned earlier.
Hence, nonzero values of $\tau$ and $\langle L \rangle$ are expected 
to yield a quantum phase shift in the wavefunction $\psi(q_0)$, 
which originates from the helical atomic configuration in the ring. 
If the ring has non-uniform cross section, equation (\ref{eq:1dsyure2}) 
contains a spatially dependent scalar potential that stems from 
the geometric curvature of the cylindrical surface of the ring 
\cite{taira,shima,ono,Atanasov}. 
This potential, nevertheless, requires no qualitative revision 
in the conclusion of this paper. 

To obtain a finite $\langle L \rangle$, we suppose an external current 
$I_{\rm ext}$ that penetrates 
through the center of the ring as shown in figure \ref{fig:fig1}.
The operator $L$ is then given by 
$L=-i \hbar \partial/\partial \theta -eB\left(q_1^2+q_2^2\right)/2$, where 
$\theta$ is the angular coordinate in the cross section, 
$B=\mu_0 I_{\rm ext}/\ell, \hspace{1.0mm} \ell=2\pi R_1$ and $\mu_0$ 
is the permeability of vacuum. 
We also assume a constant $\tau$ throughout the ring 
and a parabolic potential well 
$V(q_1, q_2) = m^* \omega_p^2 \left(q_1^2+q_2^2 \right) /2$ 
that strongly confines the transverse motions of electrons within the 
cross section of radius $R_2 \ll R_1$; the parameter
$\omega_p$ determines the steepness of the potential. 
These assumptions allow to write eigenfunctions of 
equation (\ref{eq:1dsyure2}) as \cite{taira2} 
\begin{eqnarray}
\psi(q_0) = \psi_{\rm unt}(q_0) \exp\left(-\frac{i \tau}{\hbar} \int_0^{q_0}
\langle L \rangle dq_0'\right), 
\label{eq:solution}
\end{eqnarray}
where $\psi_{\rm unt}$ is an eigenfunction 
for an untwisted quantum ring (i.e., $\tau = 0$). The ground-state 
function $u_g(q_1, q_2)$ in the cross section reads \cite{Fock,Darwin} 
\begin{eqnarray}
u_g(q_1,q_2) = \displaystyle\frac{1}{\sqrt{\pi}\ell_{\Omega}}\exp\left(-\frac{q_1^2+q_2^2}{2\ell_{\Omega}}\right), 
\label{eq:solution2}
\end{eqnarray}
where $\ell_{\Omega}=\sqrt{\hbar/(m^* \Omega)}$, 
$\Omega=\sqrt{\omega_p^2+(\omega_c/2)^2}$ and $\omega_c=eB/m^*$ is the 
cyclotron frequency. 
Then, we can prove that
\begin{eqnarray}
\frac{\langle L \rangle}{\hbar} =-\frac{I_{\rm ext}}{I_0}\frac{1}{\sqrt{4+\displaystyle\left(\frac{I_{\rm ext}}{I_0}\right)^2}}\ , \ I_0=\frac{m^* \omega_p \ell}{e \mu_0}.
\label{eq:expectation} 
\end{eqnarray}
Equations (\ref{eq:solution}) and (\ref{eq:expectation}) state that 
nonzero $I_{\rm ext}$ gives rise to quantum phase shift by 
$\tau \langle L \rangle \ell/\hbar$ in the eigenstate $\psi(q_0)$, 
in which the magnitude of the shift is determined by the formula 
(\ref{eq:expectation}). 

\begin{figure}
\begin{center}
\includegraphics[width=6.8cm]{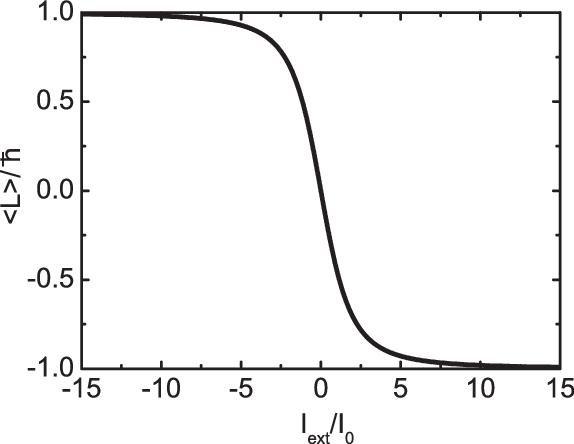}
\caption{\label{fig:fig2} 
Expectation value 
$\langle L \rangle$ of the cross-sectional angular momentum operator 
$L$ of the twisted quantum ring. 
It decreases monotonically in response to an increase in the external 
current $I_{\rm {ext}}$ that penetrates within the ring 
(see figure \ref{fig:fig1}). 
}
\end{center}
\end{figure}

Figure \ref{fig:fig2} shows how $\langle L \rangle$ depends on 
the external current $I_{\rm ext}$ normalized by $I_0$. 
$\langle L \rangle$ decreases monotonically with increasing $I_{\rm ext}$, 
while it converges to $-\hbar$ (or $+\hbar$) in the limit of 
$I_{\rm ext} \rightarrow + \infty$ (or $- \infty$). 
Furthermore, the decay in $\langle L \rangle$ is steep only within the 
region $-5 < I_{\rm ext}/I_0 < 5$. 
These features of $\langle L \rangle$ imply that the torsion-induced 
phase shift $\tau \langle L \rangle \ell/\hbar$ shows a significant 
response to the change in $I_{\rm ext}$ when $I_{\rm ext}$ lies 
within the region above. 
The flux-free interference effect in a helical circuit is a direct 
consequence of the torsion-induced phase shift, as demonstrated later. 
\section{Electron interferometer with a twisted quantum ring}
Let us consider the conductance of a twisted-ring based interferometer 
measured by two terminals, as illustrated in figure \ref{fig:fig1}.
Two different paths connecting the two points P and Q have
the same length of $\pi R_1$. 
The electron's path along two semi-infinite leads and the two 
semicircular arcs is parametrized by $x$ 
such that $x=0$ at P 
and $x=\pi R_1$ at Q; electron flow inserted from $x=-\infty$ bifurcates 
at P, passing through either of the two branches (i.e., semicircles 
of the ring) until getting confluent at Q, and then flow away 
toward $x=+\infty$.

The wave functions $\psi_s(s=1,2,3,4)$ corresponding to the four 
different regions depicted in figure \ref{fig:fig1} are given by 
\cite{ohe}
\begin{eqnarray}
       \begin{array}{l}
\psi_1(x) = A_1e^{ikx}+B_1e^{-ikx}, \\
\vspace{1.0mm}
\psi_2(x) = A_2e^{i(k+a)x} + B_2e^{-i(k-a)x}, \\
\vspace{1.0mm}
\psi_3(x) = A_3e^{i(k-a)x} + B_3e^{-i(k+a)x},  \\
\vspace{1.0mm}
\psi_4(x) = A_4e^{ikx}.
       \end{array}
\label{eq:wavefunction}
\end{eqnarray}
Here, $k$ is the wavenumber of the incident electron, 
and $a \equiv \tau \langle L \rangle/\hbar$ is a wavenumber shift
caused by $I_{{\rm ext}}$. 

The wavefunctions $\psi_s$ satisfy the connection conditions:
$\psi_1=\psi_2=\psi_3$ and $\partial_x \psi_1=\partial_x \psi_2+
\partial_x \psi_3$ at P, and $\psi_4=\psi_2=\psi_3$,
$\partial_x \psi_4=\partial_x \psi_2+\partial_x \psi_3$ at Q.
Applying the conditions to equation (\ref{eq:wavefunction}), we obtain 
the conductance $G$ of the system by using the two-terminal Landauer 
formula \cite{landauer}, 
\begin{eqnarray}
G=\frac{2e^2}{h}|A_4|^2 =\frac{2e^2}{h} \big|(1+\theta_1)A_2+2\theta_1(A_3-1)\big|^2,
\label{eq:transprob} 
\end{eqnarray}
where
\begin{eqnarray}
A_2 = \frac{\alpha'\gamma-\beta\gamma'}{\alpha\alpha'-\beta\beta'}, \
A_3 = -\frac{\beta'}{\alpha'}A_2+\frac{\gamma'}{\alpha'}, 
\label{eq:D2D3} 
\end{eqnarray}
and 
\begin{eqnarray}
       \begin{array}{l}
\alpha(k,a) =k\left[2(2k+a)-(k+a)(\theta_2/\theta_1)+(k-a)\theta_2\right], \\
\vspace{1.0mm}
\beta(k,a) =k\left[2k+a+(a/\theta_1)+2(k-a)\theta_2\right], \\
\vspace{1.0mm}
\gamma(k,a) =2k[2k+a+(k-a)\theta_2],  \\
       \end{array}
\label{eq:abc} 
\end{eqnarray}
with the notations 
$\theta_1=\exp(-ik\ell), \theta_2=\exp(ia\ell)$
and $\xi'(k,a)=\xi(k,-a)$ for $\xi=\alpha,\beta,\gamma$. 
Once determining the dimensionless parameters $k \ell$ and $a \ell$ 
(or equivalently, $k\ell$, $\tau\ell$ and $\langle L \rangle/\hbar$), 
we can evaluate $G$ by using equations (\ref{eq:transprob})-(\ref{eq:abc}). 
If we impose $a=0$ in equation (\ref{eq:abc}), the expression of $G$ reduces to that of 
an ordinary un-twisted interferometer \cite{ohe} 
\begin{eqnarray}
G=\frac{2e^2}{h} \frac{32}{41-9 \cos(k\ell)}.
\label{eq:transprobnomagnet} 
\end{eqnarray}
To make concise arguments, we omit the electron's spin-dependent transport 
\cite{jpcm3} nor impurities/structural disorder \cite{jpcm4} in the system, 
though each of them is expected to yield interesting consequences 
similarly to the case of un-twisted systems. 

\section{Results}
\begin{figure}
\begin{center}
\includegraphics[width=8.0cm]{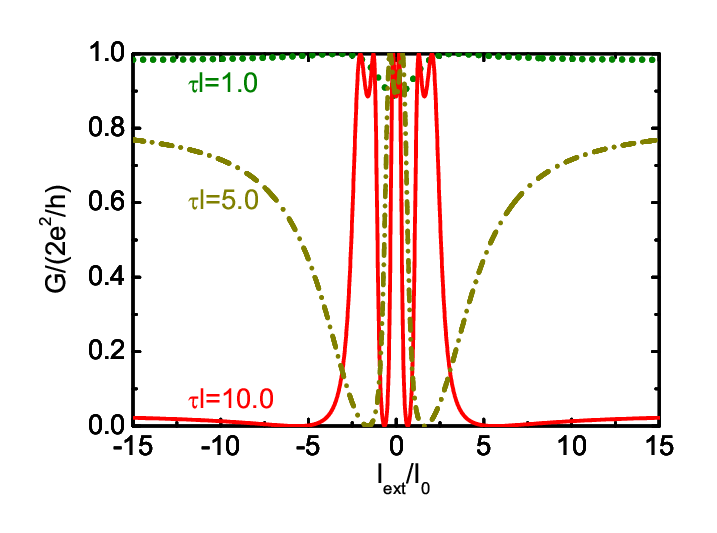}
\caption{\label{fig:fig3a} 
Dimensionless conductance
$G/(2e^2/h)$ as a function of the normalized external current 
$I_{\rm ext}/I_0$. 
The parameter $k\ell =1.0$ is fixed for all curves that are each 
associated with different values of $\tau \ell$. 
An oscillation of $G$ is observed around 
$I_{\rm ext} =0$ at $\tau\ell=5.0,\ 10.0$, and larger $\tau\ell$ (not shown).
}
\end{center}
\begin{center}
\hspace{0.1truecm}
\includegraphics[width=8.0cm]{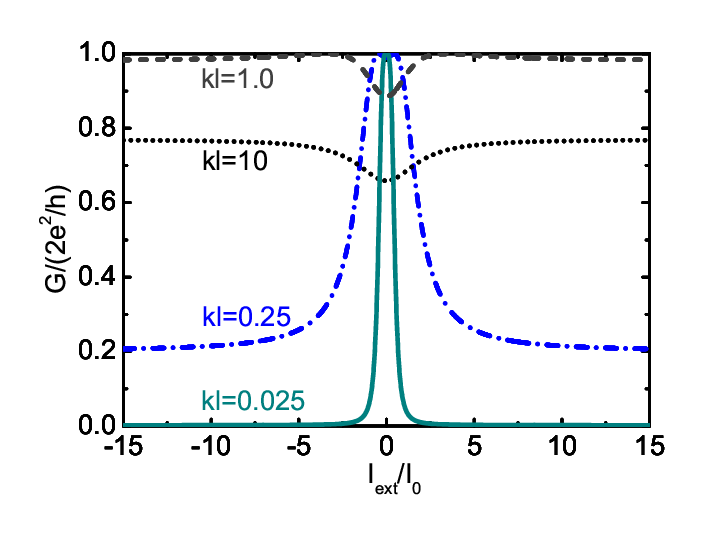}
\caption{\label{fig:fig3b} 
$I_{\rm ext}$-dependence of 
$G/(2e^2/h)$ with $\tau \ell =1.0$ being fixed. 
As $k\ell$ increases, the width of the upward peak that initially arises at 
$I_{\rm ext} =0$ becomes broader, and finally a slight hollow emerges 
at $I_{\rm ext} =0$. 
}
\end{center}
\end{figure}
%
Figures 3 and 4 show the plots of the dimensionless conductance 
$\tilde{G} \equiv G/(2e^2/h)$ as a function of the dimensionless current 
$\tilde{I}_{\rm ext} \equiv I_{\rm ext}/I_0$ 
for various values of $\tau \ell$ and $k \ell$. 
In figure \ref{fig:fig3a} (or figure \ref{fig:fig3b}), 
we fix $k\ell=1.0$ (or $\tau\ell=1.0$) and choose several values of 
$\tau\ell$ (or $k\ell$) as indicated. 
In both figures, the curves of $\tilde{G}$ exhibit waveforms within 
the region $\left| \tilde{I}_{\rm ext} \right| < 5$ but smooth 
(or almost constant) behaviors outside the region: The two contrasting 
features of $\tilde{G}$ in the two regions are attributed to the nonlinear 
response in $\langle L \rangle$ to $I_{\rm ext}$, which will be discussed soon later. 

The most important observation in figure \ref{fig:fig3a} is an 
oscillation in $\tilde{G}$ for large $\tau \ell$, 
which does not take place in figure \ref{fig:fig3b}. 
This conductance oscillation is what we call the flux-free 
interference effect, 
the peculiar phenomenon to the twisted-ring based AB interferometer.
The magnitude of oscillation can be enhanced 
(i.e., the oscillation period is shortened) by increasing 
$\tau \ell$, since the larger $\tau \ell$ results in the larger 
torsion-induced phase shift $\tau\langle L \rangle \ell / \hbar$. 
What value of $\tau \ell$ should be required for observing the 
flux-free conductance oscillation strongly 
depends on the condition of $k\ell$ we choose. 
When $k\ell=1.0$, for instance, the oscillation disappears for 
$\tau \ell \ll 10.0$ as shown in figure \ref{fig:fig3a}. 
An exhaustive study covering wide ranges of the parameter space 
makes clear the required values of $\tau\ell$ 
and $k\ell$, which will be given elsewhere. 

It is also noteworthy that the conductance oscillation 
in the present system is not periodic 
against the 
variation in $I_{\rm ext}$, which differs from 
the ordinary AB effect driven by penetrating magnetic flux. 
The non-periodic character is a consequence of the nonlinear dependence of $\langle L \rangle$ on $I_{\rm ext}$ 
(see equation (\ref{eq:expectation}) and figure \ref{fig:fig2}).
We have seen from figure \ref{fig:fig2} that
 $\langle L \rangle$ decreases steeply with increasing $I_{\rm ext}$ around $I_{\rm ext} \sim 0$ (at $|I_{\rm ext}| \gg 0$) and it decreases gently at the regions of $|I_{\rm ext}| \gg 0$. 
This means that the phase shift $\tau \langle L \rangle \ell/\hbar$ shows sensitive (insensitive) response to a change in $I_{\rm ext}$ at $I_{\rm ext} \sim 0$ ($|I_{\rm ext}| \gg 0$), and thus it reaches $2\pi$ by only a small (very large) increase in $I_{\rm ext}$ at $I_{\rm ext} \sim 0$ ($|I_{\rm ext}| \gg 0$). As a result, the conductance oscillates densely (sparsely) at $I_{\rm ext} \sim 0$ ($|I_{\rm ext}| \gg 0$) since the oscillation stems from the quantum interference caused by the phase shift $\tau \langle L \rangle \ell/\hbar$. 

As a by-product, we show in figure \ref{fig:fig3b} the 
$I_{\rm ext}$-dependence of $G$ for a fixed $\tau \ell$  ($\tau \ell=1.0$) 
and various $k\ell$. 
At $k\ell \ll 1.0$, a sharp peak arises at $I_{\rm ext}=0$ 
whose peak width broadens gradually with increasing $k\ell$. 
The peak height at $I_{\rm ext}=0$ is given by 
equation (\ref{eq:transprobnomagnet}), thus it oscillates with increasing $k\ell$. 
When $k\ell$ exceeds $1.0$, then $G$ becomes almost constant 
over the range of $I_{\rm ext}$ we have considered, 
and it shows a slight hollow at $I_{\rm ext}=0$. 
The almost constant behavior of $G(I_{\rm ext})$ indicates that the 
torsion-induced phase shift gives little contribution to the 
motion of electrons whose energies are large enough to satisfy 
$k\ell \geq 1.0$. 

\section{Discussions}
We remark that the numerical results in figures \ref{fig:fig3a} and 
\ref{fig:fig3b} are based on the presence of $I_{\rm ext}$ that 
threads the center of the ring. 
An experimental realization of such the setup may not be feasible 
in a straightforward manner, 
since the ring radius $R_1$ should be small enough to keep 
the quantum coherence of mobile electrons. 
Still, we can build an equivalent setup to the above by 
applying an external magnetic field $B_{\rm ext}$, instead of $I_{\rm ext}$, 
to a portion of the ring in a {\it tangential} direction. 
This is because the tangential field $B_{\rm ext}$ engenders nonzero 
$\langle L \rangle$ in the cross section and thus a phase shift 
in $\psi(q_0)$. 
A field strength of $B_0 \sim 10$T is required to observe the flux-free 
interference effect, provided the ring of the cross-sectional radius $R_2 \sim 10$nm; 
we can estimate it from the relations of 
$B_0=\mu_0 I_0/\ell=m^*\omega_p/e$ and 
$\hbar\omega_p\sim m^* \omega_p^2R_2^2/2$ (see Sec.~2). 
This field strength is accessible in the existing nanometric measurements, 
thus supporting an experimental feasibility of our results.

It is noticed that our attention has been limited to non-interacting electrons. When taking into account a Coulomb interaction between electrons, the amplitude of the conductance oscillation is expected to decrease to a degree. In fact, such the amplitude reduction caused by the Coulomb interaction was experimentally observed for un-twisted AB interferometers \cite{ee,ee2}, where sizeable interference patterns still remained to be obtained. A further interesting issue would be transport properties of the Tomonaga-Luttinger liquid (TLL) state in an AB interferometer. In one-dimensional systems, the slightest correlation between electrons can lead to the TLL states, a highly collective state of matter \cite{tll}. Recent studies have revealed that the conductance of TLL-based AB interferometers shows anomalous interference patters that are substantially different from those of the ordinary (non-interacting) AB system \cite{ee3,ee4}. These facts pose a question as to what happens in the helical AB interferometer composed of the TLL states, which requires establishing newly the bosonization theory for a twisted quantum wire. 
\section{Conclusion}
We theoretically predicted a non-trivial conductance oscillation 
in a twisted quantum ring that is free from threading magnetic flux. 
Helical atomic configuration inside the ring gives rise to a phase shift 
in the electron's eigenstates by $\tau\langle L\rangle \ell/\hbar$, 
where $\tau$ the internal torsion of the ring, $\langle L \rangle$ the 
angular momentum expectation value in the cross section, and 
$\ell$ the ring perimeter. 
The phase shift induces the flux-free conductance oscillation 
in response to 
a change in the external current $I_{\rm ext}$ or a tangential 
magnetic field $B_{\rm ext}$, in which neither $I_{\rm ext}$ nor $B_{\rm ext}$ 
yields magnetic flux threading the ring. 
Our results suggest untouched quantum nature in actual 
low-dimensional nanostructures composed of helical atomic configuration.

\ack

We would like to express our thanks to K Yakubo for illuminating 
discussions. 
One of the authors (HT) acknowledges K W Yu and M Arroyo 
for their comments and hospitalities during the stay in CUHK and 
UPC.
HT also thanks the financial supports from the Japan Society for the 
Promotion of Science for Young Scientists and the
Excellent Young Researcher Overseas Visit Program.
This work is supported by a Grant-in-Aid for 
Scientific Research from the MEXT, Japan.
Numerical calculations were performed in part by using the 
facility of Supercomputer Center, ISSP, University of Tokyo.



\end{document}